# The OPERA magnetic spectrometer

M. Ambrosio, R. Brugnera, S. Dusini, B. Dulach, C. Fanin, G. Felici, F. Dal Corso, A. Garfagnini, F. Grianti, C. Gustavino, P. Monacelli, A. Paoloni, L. Stanco, M. Spinetti, F. Terranova, L. Votano

***Abstract*— The OPERA neutrino oscillation experiment foresees the construction of two magnetized iron spectrometers located after the lead-nuclear emulsion targets. The magnet is made up of two vertical walls of rectangular cross section connected by return yokes. The particle trajectories are measured by high precision drift tubes located before and after the arms of the magnet. Moreover, the magnet steel is instrumented with Resistive Plate Chambers that ease pattern recognition and allow a calorimetric measurement of the hadronic showers. In this paper we review the construction of the spectrometers. In particular, we describe the results obtained from the magnet and RPC prototypes and the installation of the final apparatus at the Gran Sasso laboratories. We discuss the mechanical and magnetic properties of the steel and the techniques employed to calibrate the field in the bulk of the magnet. Moreover, results of the tests and issues concerning the mass production of the Resistive Plate Chambers are reported. Finally, the expected physics performance of the detector is described; estimates rely on numerical simulations and the outcome of the tests described above.**

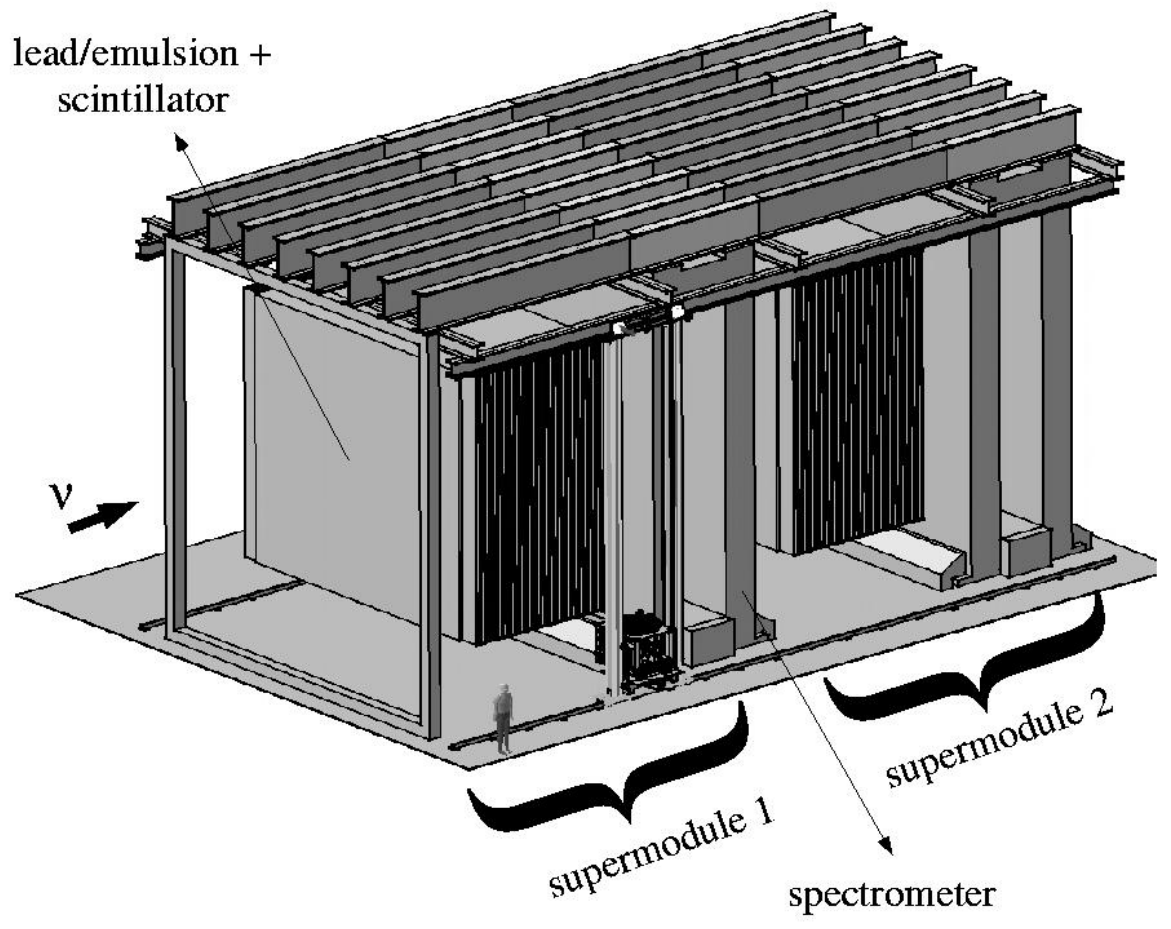

Fig. 1. The OPERA experiment at LNGS.

## I. Introduction

OPERA is a long-baseline neutrino experiment currently under construction at the Gran Sasso underground laboratories (LNGS) [1]. Its aim is the observation of $\nu_\mu \to \nu_\tau$ oscillations in the parameter region indicated by Super-Kamiokande [2] through direct observation of $\nu_\tau$ charged current interactions. The detector design is based on a massive lead/nuclear emulsion target (ECC) complemented by electronic detectors (scintillator bars) that allow the location of the event and drive the scanning of the emulsions. This instrumented target is followed by a magnetic spectrometer which measures charge and momentum of penetrating tracks. The instrumented target and the spectrometer constitutes one "supermodule". OPERA is made up of two supermodules located in the Hall C of LNGS (see Fig. 1). The magnets contribute to the kinematic reconstruction of the event performed by the ECC and the scintillators and suppress the background coming from charm production through the identification of secondary antimuons.

M. Ambrosio is with the Department of Physics, Univ. of Napoli and INFN sez. di Napoli, Napoli, Italy.

R. Brugnera, S. Dusini, C. Fanin, F. Dal Corso, A. Garfagnini, L. Stanco are with the Department of Physics, Univ. of Padova and INFN sez. di Padova, Padova, Italy.

B. Dulach, G. Felici, A. Paoloni, M. Spinetti, F. Terranova, L. Votano are with the Laboratori Nazionali di Frascati dell'INFN, Frascati (Rome), Italy.

C. Gustavino is with the Laboratori Nazionali del Gran Sasso dell'INFN, Assergi (L'Aquila), Italy.

F. Grianti is with the Department of Physics, Univ. of Urbino and INFN-LNF., Urbino, Italy.

P. Monacelli is with the Department of Physics, Univ. of L'Aquila and INFN-LNGS, L'Aquila, Italy.

The magnet (Fig. 2) is made up of two vertical walls of rectangular cross section and of top and bottom flux return path. The walls are built lining twelve iron layers (5 cm thickness) interleaved with 2 cm of air gap, allocated for the housing of the active detectors (Resistive Plate Chambers, RPC). Each iron layer is made up of seven slabs $50 \times 1250 \times 8200$ mm$^3$. These slabs are precisely milled along the two 1250 mm long sides connected to the return yokes to minimize air gaps along the magnetic circuit. The slabs are bolted together to increase the compactness and the mechanical stability of the magnet which, hence, can act as a base for the emulsion target supports. The nuts holding the bolts also serve as spacers between the slabs and fix the 20 mm air gap where the RPC are mounted. The slabs are bolted at the top and bottom edges to the return flux paths. The return yokes consist of six steel substructures (1250 mm width) and two additional 625 mm blocks (see Fig. 2). The bolts and the nuts prevent sliding of the slabs making the multilayer wall, together with the return path, a compact self-supporting structure. Moreover, this design permits us to assemble the magnet in a fast and neat way: after the positioning of the bottom return yokes and the lower coil, the installation proceeds by mounting alternately the iron layers and the inner tracker planes starting from the two innermost iron/RPC layers. Once the two outermost layers are positioned, the structure is completed and stabilized by the installation of the upper return yoke. The overall weight of the magnet is 990 ton. It is magnetized by means of two coils, 20 turns each, installed in the top and bottom flux return path. The nominal current flowing in the coils is 1600 A, corresponding to an overall magnetomotive force of 64000 A·turns. The average field expected along the walls is 1.57 T with an uniformity along the height better than 5%.

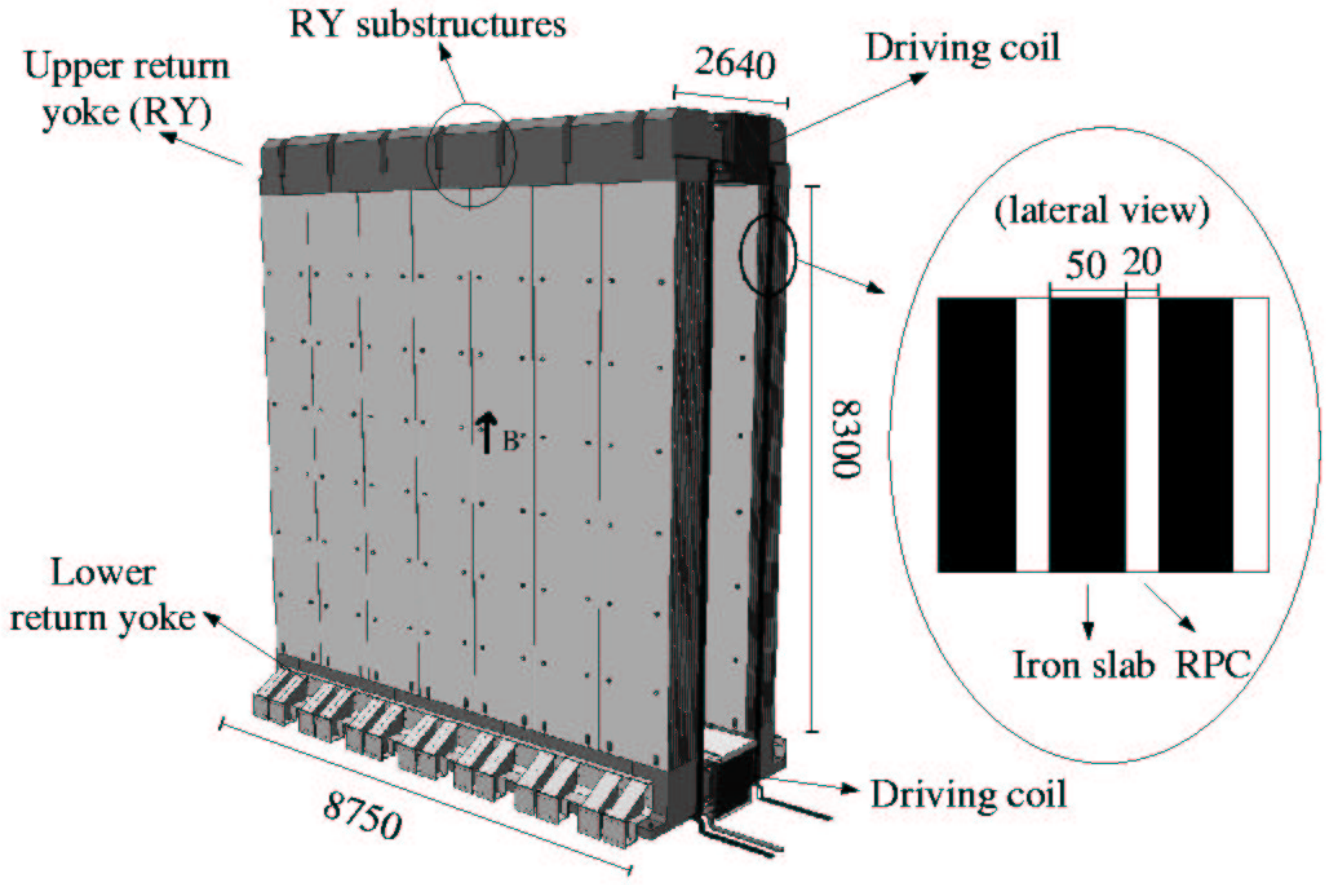

Fig. 2. 3D view of the OPERA magnet. Units are in mm.

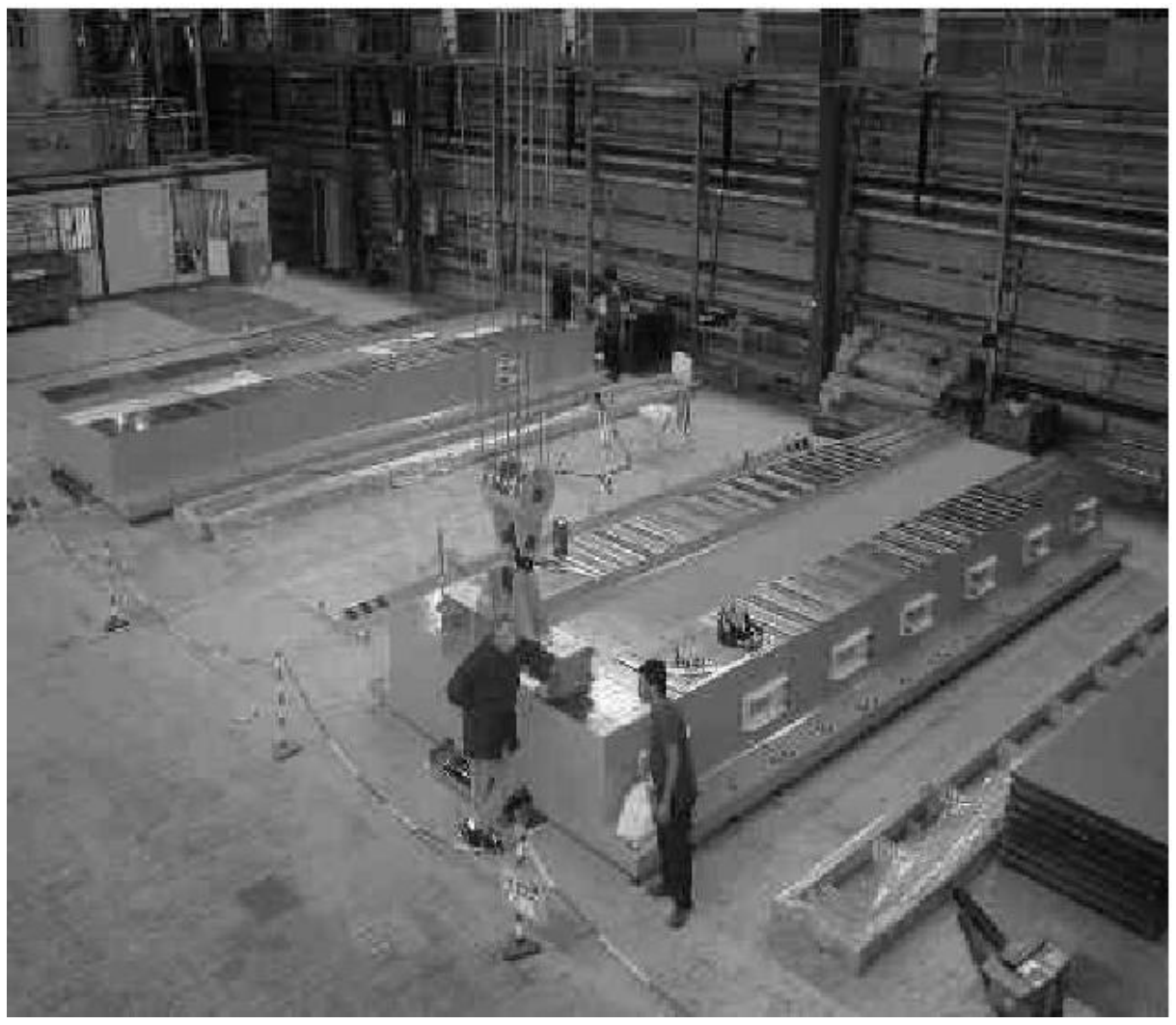

Fig. 4. The Hall C of the Gran Sasso laboratories after the installation of the bottom return yokes (Sep. 2003).

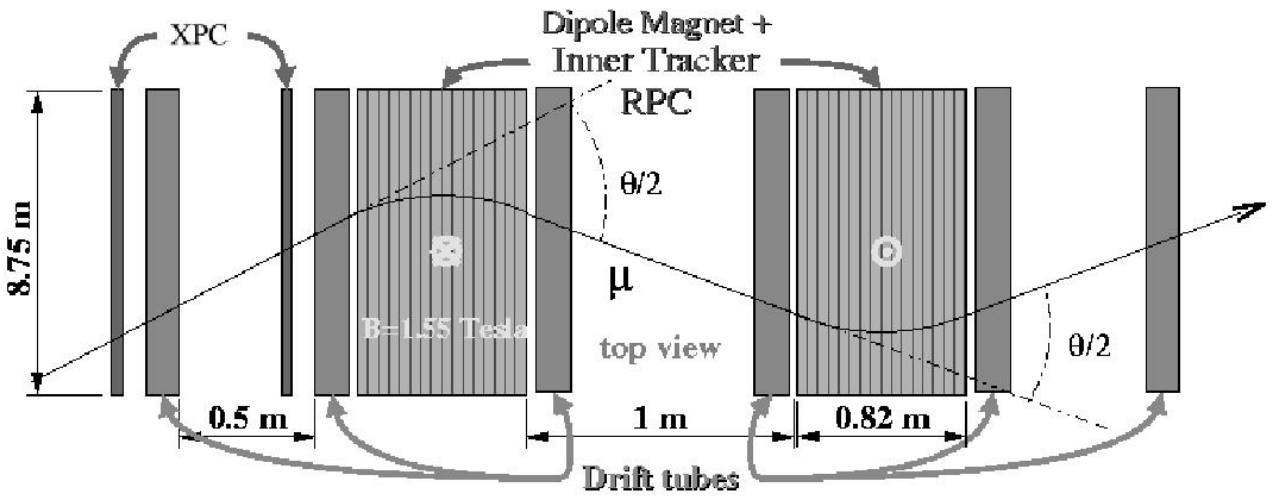

Fig. 3. The OPERA muon system. A track trajectory along the drift tube chambers, the XPC and the two walls of the instrumented magnet is also shown ($dE/dx$ losses are neglected).

The particle trajectories are measured by layers of vertical drift tube planes located before and after the walls (Fig. 3). Left-right ambiguities are resolved by the two-dimensional measurements of the spectrometer RPC and additional RPC located between the ECC target and the first magnet wall with $\pm 45^o$ orientation pickup strips ("XPC")[1]. Moreover, the RPC allow a coarse tracking inside the magnet to identify muons, perform pattern recognition and ease track matching between the precision trackers [6]. They also provide a measurement of the tail of the hadronic energy leaking from the target and of the range of muons which stop in the iron.

[1]Details on the design and construction of the OPERA drift tubes and XPC can be found in [1], [3], [4], [5].

The construction of OPERA started in March 2003. The installation of the bottom return yokes and coils has been carried out (Fig. 4). Mounting of the vertical slabs started in November 2003, while the completion of the first spectrometer and the magnetization of the steel for cosmic ray runs is planned for summer 2004.

## II. Iron properties

The magnet is the basic support structure of OPERA. During the mounting of the walls the mechanical structure of the spectrometers undergoes significant stresses. Moreover, after the completion of the installation, the magnet is sensitive to the weight of the ECC target, the magnetic forces and possible seismic stresses. Hence, severe requirements have been applied to the mechanical properties of the steel, especially for what concerns its breaking strength, elongation and yield strength. On the other hand, the steel must be appropriate for magnetic applications (high magnetic permeability) in order to achieve the nominal field for deflection of charged particles. A S235 JR steel (unalloyed steel for magnetic application) in compliance with UNI EN 10025 has been chosen. In addition, upper limits on the weight fractions for Carbon, Phosphorus and Sulphur have been specified. The weight fraction of Manganese affects significantly the mechanical properties of the steel. Hence, an higher fraction of Mn has been allowed for the slab steel. It results, however, in a deterioration of the magnetic properties (see below). On the other hand, looser mechanical requirements apply on the return yokes and a significantly lower weight fraction of C and Mn could be allowed. Boron contamination is kept below 5 ppm. The slabs have been produced by DUFERCO (Clabecq, Belgium). Machining has been carried out by MELONI (Tivoli, Italy). The return yokes have been produced and machined by FOMAS (Osnago, Italy). The chemical analysis is done by the steel producers for each production batch. Moreover, the magnetic properties of the steel produced for the iron walls and return yokes are checked in a direct manner: small toroidal samples have been prepared for each batch; hysteresis curves have been drawn for every sample at different maximum values of $H$ and the corresponding coercivity has been determined. A systematic difference between the properties of the steel used for the vertical slabs and the one used to construct the prototype of the spectrometer [7] has been observed. It has been traced back to the higher Manganese fraction allowed to achieve the mechanical specifications. The relative magnetic permeabilities ($\mu_r \equiv \mu/\mu_0$) are shown in Fig. 5 (lower plot) together with the $B-H$ curves (upper plot) for the prototype steel (full dots) and one batch of slab steel (empty dots). We expect a difference in the magnetic field $B$ of about 3% at the nominal magnetomotive force.

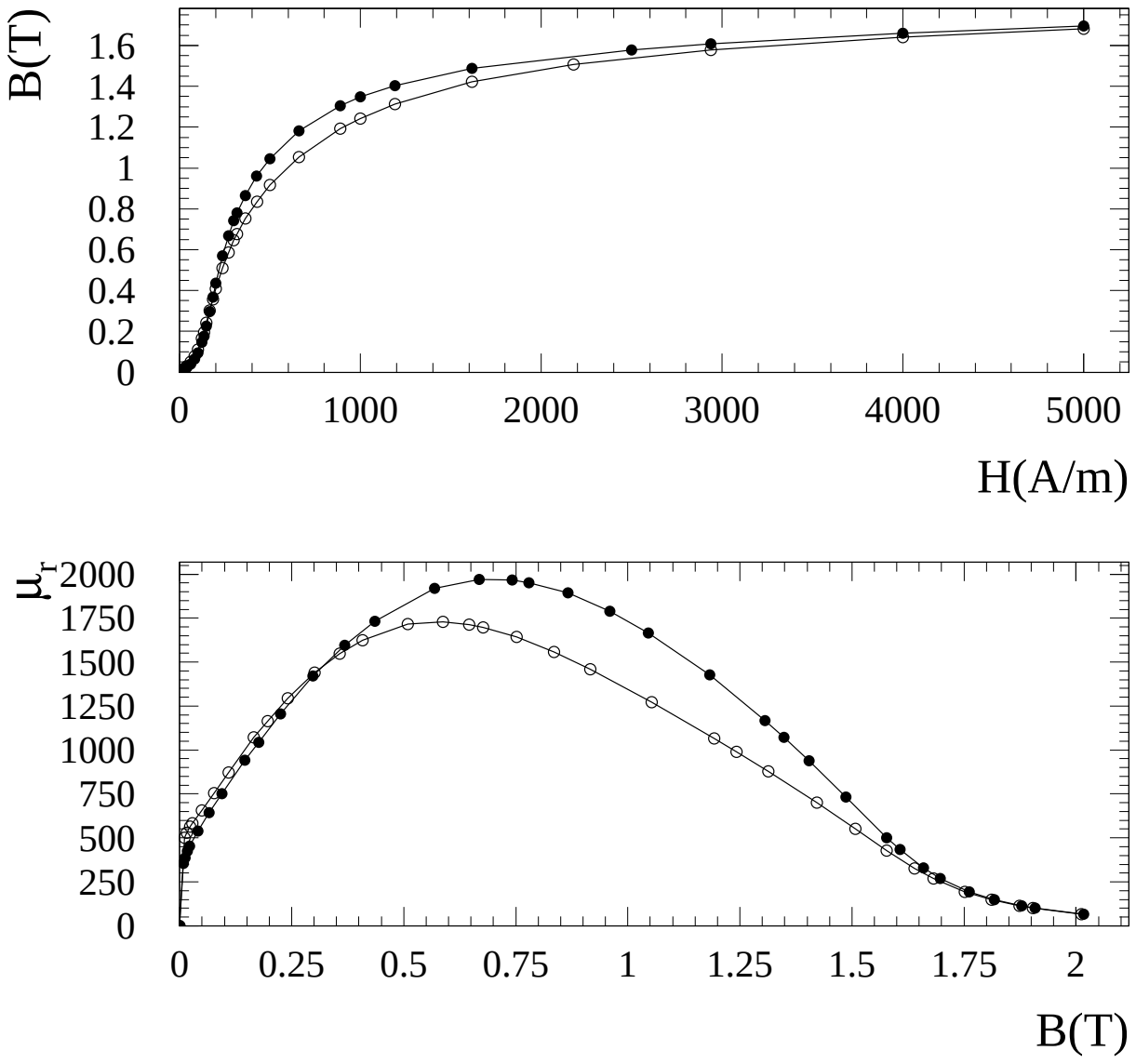

Fig. 5. Comparison of the magnetic properties of the steel used for the prototype (full dots) with one heat of the steel produced for the OPERA magnet (empty dots). Upper plot: $B$ versus $H$. Lower plot: Relative magnetic permeabilities ($\mu_r \equiv \mu/\mu_0$) versus $B$.

## III. MONITOR OF THE MAGNETIC FIELD

The absolute measurement of the magnetic flux $B$ is based on the voltage induced in a set of pickup coils during the ramp-up (ramp-down) of the power supply ("ballistic measurement"). The variation of the current flowing in the driving coils induces a change in the magnetic flux cut by the pickup coil; the induced voltage is

$$V = -\frac{d\Phi(t)}{dt} = -\frac{d}{dt}\int_S \vec{B}(t)\cdot\vec{n}\ ds = -SN\frac{d}{dt}\langle B\rangle(t) \quad (1)$$

and integrating:

$$\frac{-1}{SN}\int_0^t dt' V(t') = \frac{1}{SN}(\Phi(t)-\Phi(0)) = \Delta\langle B\rangle(t) \quad (2)$$

where $S$ is the cross sectional area of one turn of the pickup coil, $N$ is the number of turns and $\Delta\langle B\rangle$ is the increase of magnetic field going from $t'=0$ to $t'=t$ averaged in each point of the surface of the turn $S$. The absolute value of the field $B$ at the nominal current averaged on the coil surface can be determined after a whole hysteresis loop is completed. Three ("external") pick-up coils will be installed along the height of each magnet wall. These coils average the magnetic flux of all the 12 layers. The presence of layers with anomalous reluctances (air gaps at the connection with the return yokes) can be measured by additional ("internal") coils inserted parallel to the RPC at about half the height of the magnet. The coils will be made up of 32-wire flat cables connected through a passive board. The board allows to vary the number of turns forming the coil and includes a low-pass filter. The two ends of the coil are sent to the integrator. The induced voltage depends on $SN$ and the current gradient $di/dt$. However, the shape of the function $V(t)$ is driven by the strong nonlinear behavior of ferromagnetic materials and by the magnetic skin effect [8]. The maximum voltage during the nominal hysteresis curve is expected to be $V_{max} \simeq 1.5$ V for $N = 10$ external coils, $S = 5.25$ m$^2$ and $di/dt = 10$ A/s.

The monitoring system has been validated in a full-height prototype built in Frascati in 2001 [7] . The results are shown in Fig. 6 and are compared with the expectation from finite-element simulation based on the TOSCA [9] code. Similarly, tests of uniformity of the field have been carried out by means of mobile coils. The main sources of error are related with the integration of $V$. In particular, baseline fluctuations and environmental background represent a non negligible contribution since the integration is performed for time intervals of the order of several tenths of seconds. For the present setup, they correspond to an uncertainty of 0.03 T in the magnetic field. Systematics have been investigated performing measurements with different $di/dt$ and with $di/dt = 0$ ("empty measurements" for signal to noise evaluation). No significant biases were found. On the other hand, a systematic bias of about 4% is observed with respect to simulation. This is mainly due to the difference of magnetic properties of the steel after machining compared with the nominal sample. Moreover, additional air gaps coming from the non-ideal mechanical contacts between the slabs and the return yokes contribute to the observed deficit.

The pick-up coils can be used only during the ramp-up of the current. Relative variations of the field can be monitored indirectly by the power-supply current monitors. However, additional information will be provided by a set of Hall probes glued to the RPC, which measure the fringe field in the 2 cm air gap. These data will be used to validate the simulation and to monitor possible long term drifts of the field. Similarly, Hall probes will monitor the fringe field in air outside the magnet, particularly in the region where the phototubes reading the scintillator bars are located.

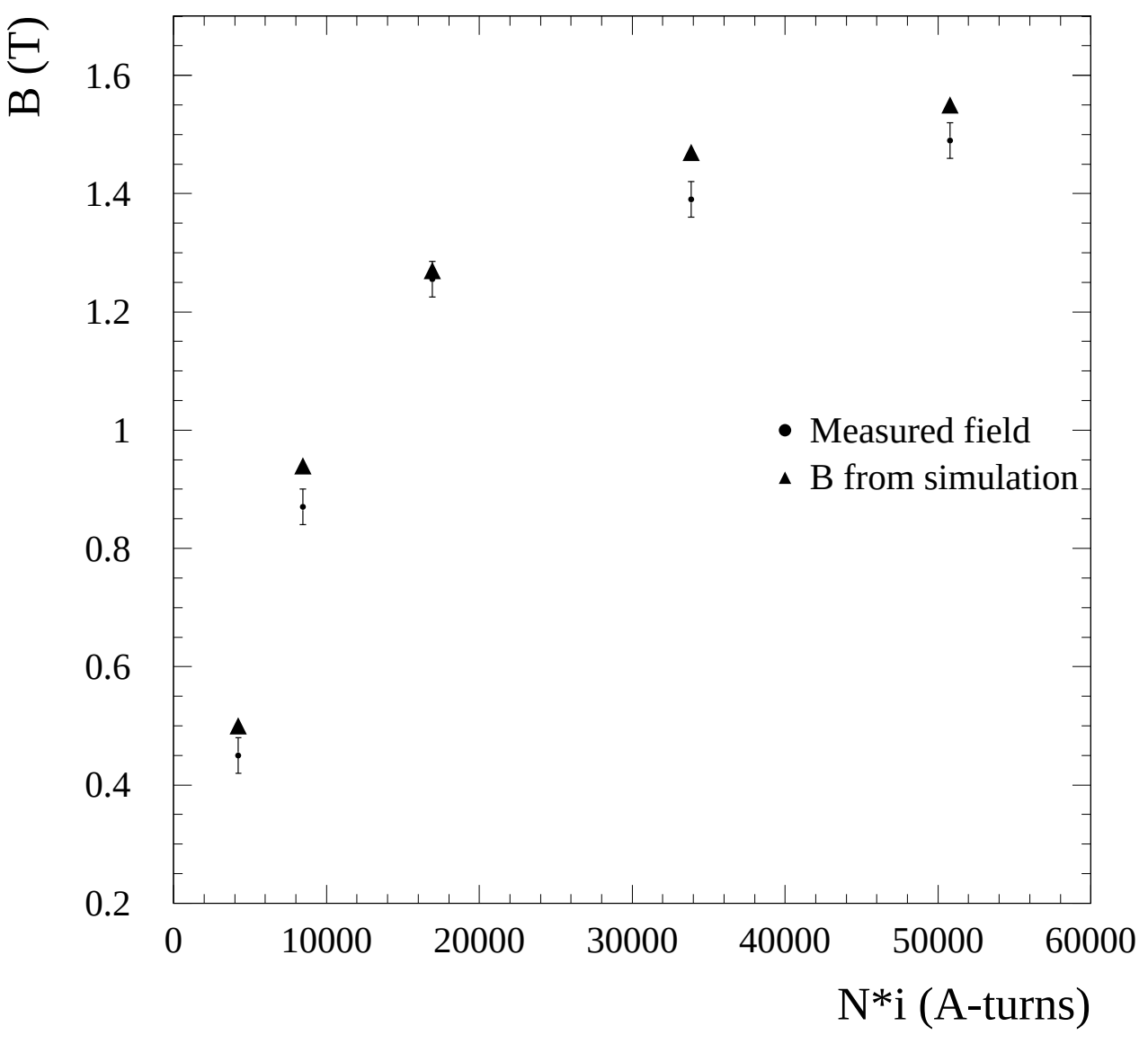

Fig. 6. Field at the pick-up coil versus current.

## IV. Resistive Plate Chambers

In the 2 cm thick gaps between the iron slabs, a layer of Resistive Plate Chambers [10] will be positioned. The $8.2\times8.75$ m$^2$ area of the magnet is covered by 21 chambers. The dimensions of the chambers are $2.91\times1.14$ m$^2$ but four different chamber types have been designed; they are shaped to optimize the sensitive area accounting for the presence of the bolts. The plates that act as electrodes are made of 2 mm thick high-pressure plastic laminate (Bakelite), with a volume resistivity of $5\ 10^{11} - 10^{12}$ $\Omega$-cm. Their external surface is painted with graphite of high surface resistivity ( 100 k$\Omega$/square) connected to the HV distribution system and covered by two 190 $\mu$m insulating films. The inner surface of the Bakelite planes is treated with linseed oil, which significantly improves the smoothness of the electrodes. Planarity of the two electrodes is assured by polycarbonate spacers (LEXAN) located on a 10 cm-square grid in the sensitive volume. The active area is filled with an argon-based gas mixture (Argon 75.5%, TFE 20%, isobutane 4%, $SF_6$ 0.5%) at atmospheric pressure. The 2-dim position measurements are provided by two orthogonal layers of copper strips. These pickup strips are not interrupted in the vicinity of the bolts but surround the dead zone as shown in Fig. 7. The chambers operate in streamer mode at a working voltage of 5.85 kV. The induced charge at the strips is of order of 100 pC, and the pulse has a rise time of 2 ns and a duration of 10 ns. Results of the prototype tests are reported elsewhere [11]. The mass production of the detectors started in Jan 2003 and the RPC for the first spectrometer have been produced. Several validation tests are in progress [12]:

- Test of gas tightness at the production site.
- Mechanical tests.
- Electrical tests.
- Efficiency measurements with cosmic muons.

The mechanical tests have been fully automatized. First of all a leakage test is carried out. The detectors are inflated at 6 mbar and temporal variations of pressure are monitored. Chambers that exhibit a leakage rate larger than $50$ $\mu$bar/min are rejected. Afterward, a mobile bridge housing a set of pistons (Fig. 8) is moved in the proximity of the LEXAN spacers. The pistons are lowered in order to exert pressure onto the upper RPC surface. An abnormal variation of pressure indicates the presence of unglued spacers. A further test determines the exact position of the defective buttons. The fraction of accepted RPC after the test is about 90%. The electrical tests have been designed to measure the current-voltage characteristics in pure argon, the current-voltage characteristics in gas mixture and the short term behavior of the RPC at voltage. The facility developed in 2003 allows to test 48 RPC simultaneously. The current-voltage characteristics in argon allows to check the cleanliness of the internal surface of the gap and to measure the average resistivity of the two electrodes. The chambers are characterized by their current derivative $dI/dV$, their primer voltage at fixed currents (I=100 nA and 500 nA) and their current at nominal operating voltage. Finally, a short term test is performed (for minimum 24 hours up to few days). The RPC are kept at fixed voltage and the time stability of the current is monitored. In particular, chambers with $dI/dV$ in Argon higher than 20 nA/kV or high currents ($>$700 nA) at the nominal operating voltage with the standard gas mixture are rejected. The cosmic ray test [13] is carried out employing a test facility built at the beginning of 2003. Two large planes of glass RPC are used for triggering purpose. The RPC under tests are placed inside boxes, (12 RPC/box), in vertical position and their efficiency are computed selecting isolated cosmic muons. The overall fraction of accepted RPC is $\sim$83%. The main source of rejection is due to the poor gluing of the spacers: about 10% of the RPC fail the mechanical test. Failure at the electric test contribute to 0.5%; poor ($<$95%) or strongly non-uniform efficiencies are observed in 6.5% of the RPC (failure of the cosmic ray test).

## V. Expected physics performances

The tests results described in Sec. III and IV allowed a realistic tuning of the full simulation for the OPERA muon system. Due to their coarse granularity, the RPC play a minor role in the determination of the momentum for particle crossing the spectrometer. In this case, the curvature is measured by a set of four Drift Tube (DT) chambers located as in Fig. 3. Each chamber is made up of four DT planes with a staggered geometry aimed at minimizing the dead zones. The drift time resolution corresponds to an uncertainty in the drift radius of about 0.5 mm. Conversely, the RPC contribute to the measurement of the momentum from range for muons stopping in the steel. The results in term of relative precision on $1/p$ are shown in Fig. 9 for thoroughgoing muons. The continuous line is the expectation from the analytical formulas [14] that keep into account the precision of the trackers (dominated by DT) and the uncertainty due to Multiple Scattering. The error bars account for the event statistics. The momentum resolution for stopping muons (momentum from range) is shown in Fig. 10. The charge misassignment probability for thoroughgoing muons in the energy range of interest (2-20 GeV) is below 0.4%.

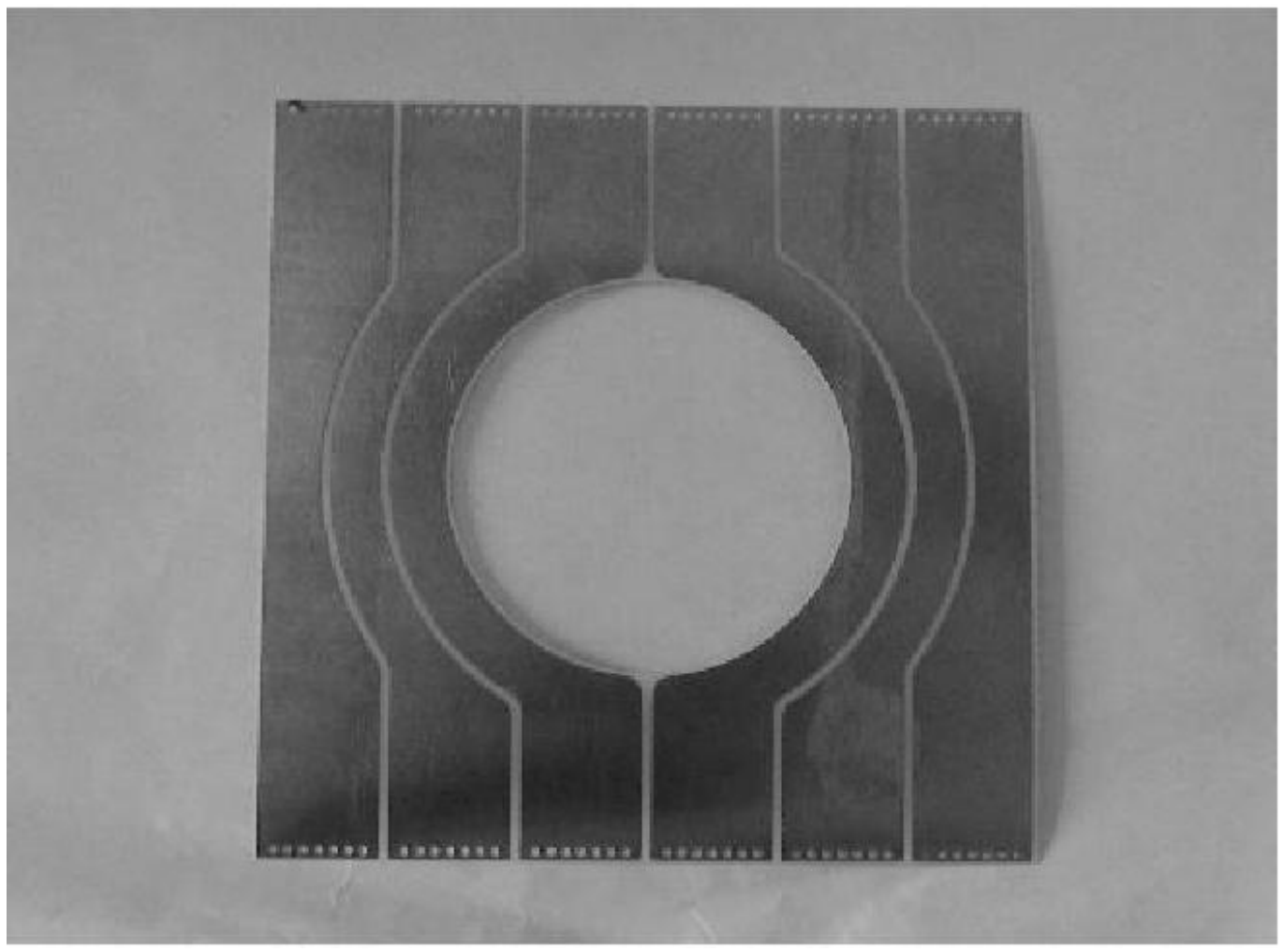

Fig. 7. Readout strips near the bolt.

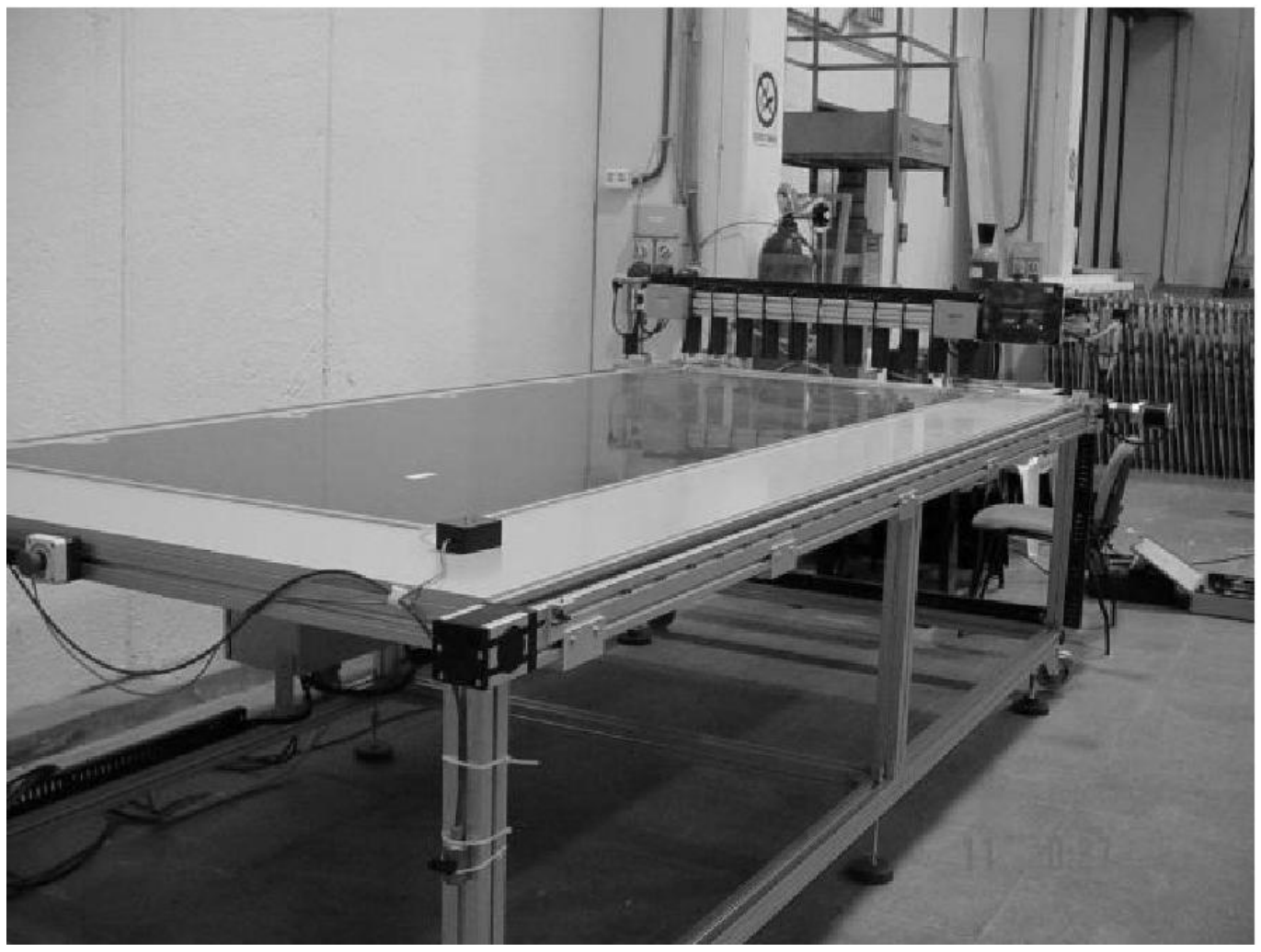

Fig. 8. Station for the mechanical tests at LNGS.

## VI. Conclusions

The design and construction of the OPERA magnetic spectrometer poses unusual engineering problems, not often encountered in the development of muon systems. The spectrometer acts both as a subdetector and as the main support structure for the emulsion-lead target. Moreover, in a deep underground area, field monitoring and calibration cannot be accomplished by the analysis of stopping cosmic muons and require a dedicated system based on pickup coils and Hall probes. Finally, the size and peculiar geometry of the spectrometer brought to the construction of large and formed Resistive Plate Chambers. In this paper we reviewed the solutions adopted in the design and construction phase. Results from the prototyping phase have been summarized and were used to validate the expected physics performance through a full simulation of the OPERA muon system.

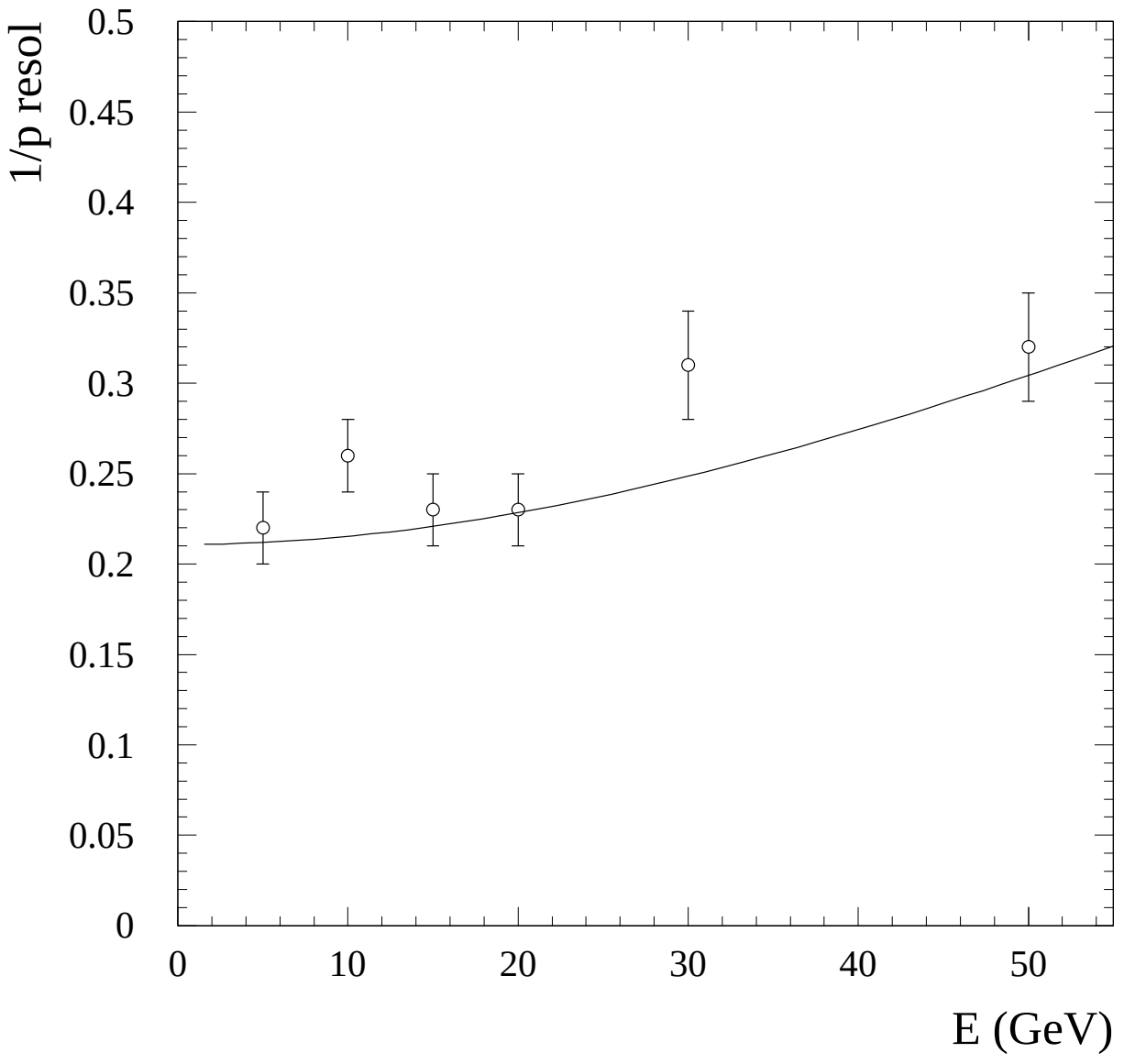

Fig. 9. Relative $1/p$ resolution for thoroughgoing muons crossing one spectrometer. The error bars account for finite MC event statistics.

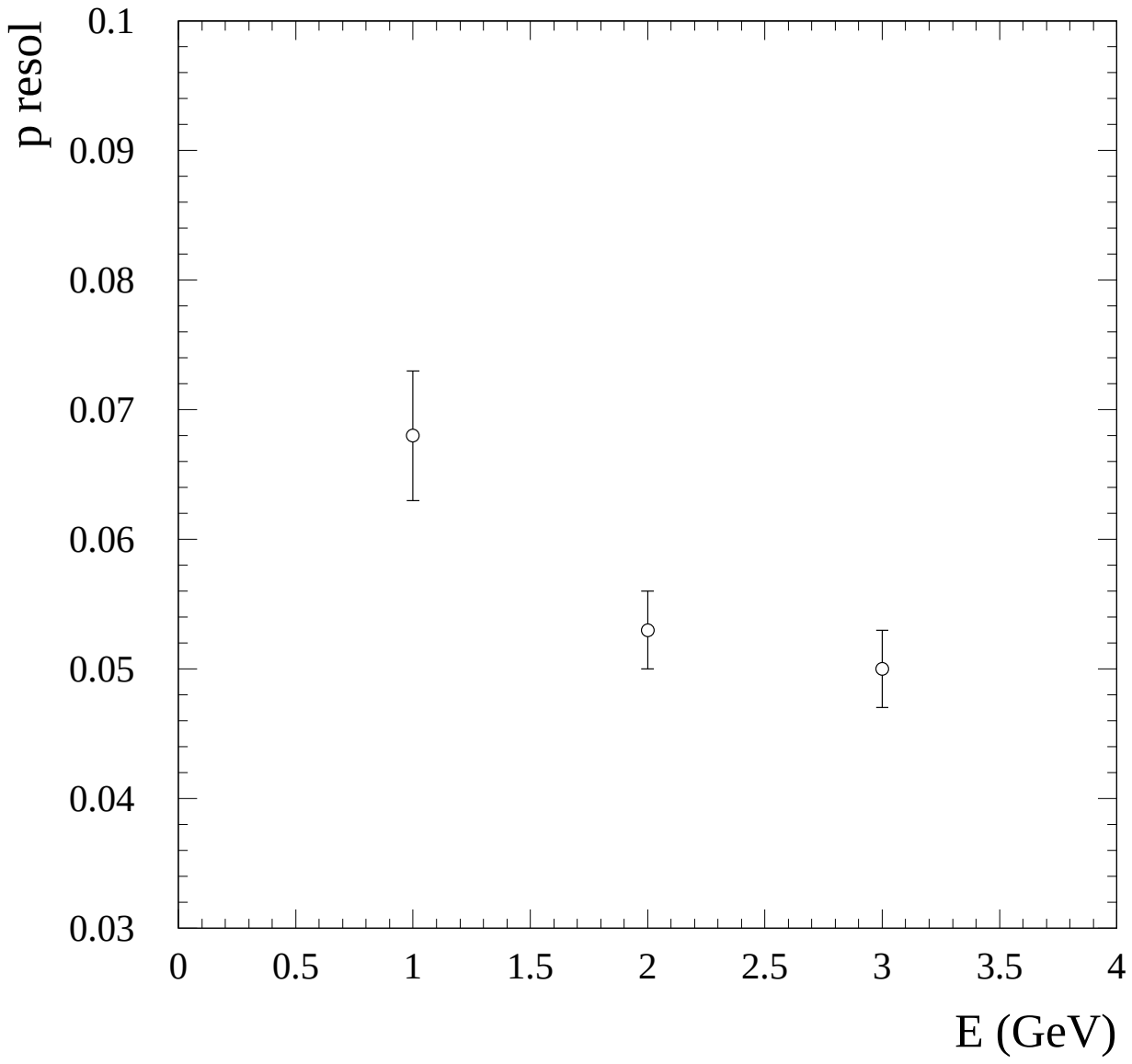

Fig. 10. Relative $1/p$ resolution for stopping muons (momentum from range). The error bars account for finite MC event statistics.

## ACKNOWLEDGMENTS

The authors would like to acknowledge the support of the SPAS and SEA groups of LNF (A. Cecchetti, G. Corradi, A. Mengucci, D. Orecchini, G. Paoluzzi, G. Papalino, M. Ventura), the LNF Accelerator Division (M. Incurvati, F. Iungo, C. Sanelli, F. Sardone) and INFN Padova (G. Barichello, E. Borsato, L. Ramina, M. Nicoletto e C. Manea).